\documentclass[3p,twocolumn]{elsarticle}

\usepackage{amsfonts}
\usepackage{amsmath}
\usepackage{algorithm}
\usepackage{algpseudocode}
\usepackage{subfig}
\usepackage{multirow}
\newcommand{\mc}[1]{\mathcal{#1}}

\DeclareGraphicsExtensions{.pdf,.png,.jpg,.eps}
\date{September 1, 2015}

\begin{document}
\begin{frontmatter}

\title{Cellular traffic offloading via Opportunistic Networking with
Reinforcement Learning}

\author[cnr]{Lorenzo Valerio\corref{cor1}} \ead{lorenzo.valerio@iit.cnr.it}
\author[cnr]{Raffaele Bruno} \ead{raffaele.bruno@iit.cnr.it}
\author[cnr]{Andrea Passarella} \ead{andrea.passarella@iit.cnr.it}
\cortext[cor1]{Corresponding author}
\address[cnr]{Institute for Informatics and Telematics, National
Research Council. Via G. Moruzzi 1, 56214 Pisa, Italy}

\begin{abstract}

  The widespread diffusion of mobile phones is triggering an
  exponential growth of mobile data traffic that is likely to cause, in the near
  future, considerable traffic overload issues even in last-generation cellular
  networks.  Offloading part of the traffic to other networks is considered a
  very promising approach and, in particular, in this paper we consider
  offloading through opportunistic networks of users' devices.  However, the
  performance of this solution strongly depends on the pattern of
  encounters between mobile nodes, which
  should therefore be taken into account when designing offloading control
  algorithms.
In this paper we propose an adaptive offloading solution based on the
Reinforcement Learning framework and we evaluate and compare the performance of
two well known learning algorithms: Actor Critic and Q-Learning.  More
precisely, in our solution the controller of the dissemination process, once
trained, is able to select a proper number
of content replicas to be injected in the opportunistic network to guarantee the
timely delivery of contents to all interested users.  We show that our system
based on Reinforcement Learning is able to automatically learn a very efficient
strategy to reduce the traffic on the cellular network, without relying on any
additional context information about the opportunistic network. Our  solution
achieves higher level of offloading with respect to other state-of-the-art approaches, in a
range of different mobility settings.  Moreover, we show that a more refined
learning solution, based on the Actor-Critic algorithm, is significantly more
efficient than a simpler solution based on Q-learning.

\end{abstract}


\end{frontmatter}

\section{Introduction}

Over the last few years we have witnessed an exponential
growth of data traffic in cellular networks.  On the one hand this is due to
the exponential spreading of  mobile devices, such as smartphones and tablets,
with multiple heterogeneous wireless interfaces. On the other hand, the
diffusion of content-centric services  among mobile users, e.g. Netflix, Youtube
or Spotify, is triggering a strong increase in the amount of traffic carried by
the cellular networks.  It has been recently estimated that  this huge
traffic growth will accelerate  in the near future~\cite{Cisco:sp}.  CISCO
forecasts that mobile data traffic  will grow at a Compound Annual Growth Rate
(CAGR) of 61\% between 2014 and 2019, reaching 15.9
Exabytes per month by
2019~\cite{Cisco:sp}, resulting in the so called \emph{data tsunami}.  The
adoption of 4G technologies is giving a significant boost to the cellular
capacity with respect to the current demands. However, projections over the next
few years show that this additional capacity may be saturated soon, as it is
foreseen to grow only by a factor of 1.4 by 2019, which may not be sufficient to
cope with the data tsunami effect. As a consequence, bandwidth crunch events
similar to those occurred with 3G networks~\cite{NYTimes:fp} may be expected, unless
alternative solutions are devised in time.


A promising  solution to reduce the burden on the cellular infrastructure is to
divert part of the traffic from the
cellular network, through
  \emph{data offloading} mechanisms~\cite{Reaz:2014aa,Talipov:2014aa, Valerio:2015aa, Asadi:2014ab}. Data offloading,
indeed, is  considered one of the most promising techniques to complement pure
cellular networks and cope with the amount of mobile data traffic expected in the
next few years~\cite{Conti:2014aa,3GPP:2013aa,3GPP:2013ab} (see
  Section~\ref{sec:related} for a brief survey on the main offloading approaches
presented in the literature).

In this paper we consider content whose delivery is delay tolerant and the
offloading is based on a device-to-device (D2D) approach. In this
context Opportunistic Networks
offer a very powerful alternative to relieve part of the network traffic from
the cellular infrastructure.  Opportunistic networks
\cite{Pelusi:2006th,Ferretti:2013aa,Mota:2014aa,Boldrini:2014ab} are self-organising mobile
networks where the existence of simultaneous end-to-end paths between nodes is
not taken for granted, while disconnections and network partitions are the rule.
Opportunistic networks support multi-hop communication by temporarily storing
messages at intermediate nodes, until the network reconfigures and better relays
(towards the final destinations) become available.  

The main research issues in Opportunistic Networks focus on the development of
analytical models of data delivery
performance~\cite{Boldrini:2014aa,Sermpezis:2014aa,Ginzboorg:2014aa}, routing
approaches that consider nodes' aggregation and privacy
~\cite{Chen:2014aa,Aviv:2014aa}, forwarding schemes for hybrid networks
(opportunistic networks combined with the
infrastructure~\cite{Mayer:2014aa}), real world implementations
~\cite{Grasic:2014aa} , applications to Vehicular
Networks~\cite{Benamar:2014aa,Kokolaki:2014aa}.

Offloading through opportunistic networks is particularly appealing when the same content is
requested by multiple users in a limited geographical area in a non-strictly
synchronised way (which rules out the possibility of multicasting on the
cellular network). Given the typical Zipf nature of content
interest~\cite{Tatar:2014aa}, this
scenario is very relevant. When offloading is used, content can be sent via the
cellular network only to a small fraction of interested users (typically
  called \emph{seeds}), while the
majority of them can be reached via opportunistic networking
techniques~\cite{Bruno:2014aa, Nordstrom:2014aa,Boldrini:2014aa, Mtibaa:2013aa}. In addition, note
that recent results also indicate that opportunistic offloading can work very well in
cooperation with cellular multicast in case of synchronised
requests~\cite{Rebecchi:2014aa}.

One of the known challenges of pure opportunistic networks is that it is very
difficult to guarantee maximum content delivery deadlines, because of the random
nature of contacts between nodes (enabled by users' mobility) on which they are
based. 
Even in case of delay-tolerant traffic, it is realistic to assume that
content would need to be delivered within a maximum deadline. This is, for
example, supported by recent studies of Web traces, which show that the value of
requested Web content drastically diminishes if not received within a certain time from the request~\cite{Tatar:2014aa}. 

In this paper (as in our previous work \cite{Valerio:2014yq} and in several
previous papers \cite{Rebecchi:2014ai,Whitbeck:2012sf}), we consider an
operator-assisted offloading system. Specifically a central  dissemination controller
decides dynamically over time to which nodes content must be sent through the
cellular network, based on the current status of the dissemination process. This
is tracked through ACKs sent by nodes (over the cellular network) upon receiving
content (over the opportunistic network). Injections of content to specific
nodes through the cellular network helps to boost the dissemination process over
the opportunistic network, and can also be used as a last-resort option to send
content to nodes that have not yet received it through the opportunistic network
by the delivery deadline.  As we show in the paper, using such an approach
guarantees maximum delivery deadlines, and reduces significantly the traffic
carried by the cellular network at the same time.

In this paper we design a new solution to control the content dissemination
process described before.  Precisely, the controller has to cope with two
problems: i) identifying how many nodes  should be delegated to disseminate a
piece of content and  ii) which of them are the most useful to speed up the
dissemination process.  As far as point i) is concerned, due to high dynamicity
of the opportunistic dissemination process, it would be desirable that the
central dissemination controller would be able to cope with this problem
autonomously, in order to reduce as much as possible planning and fine
tuning of algorithms' parameters, or previous knowledge about the behaviour of
the network nodes.
As explained in more detail in the following, we address the first problem through a Reinforcement Learning (RL)
approach.  As far as point ii) is concerned, we adopt a heuristic mechanism that
permits to identify, online, what nodes are the most useful to spread contents.

We cast the offloading problem as a RL problem, in a very general way such that
we can apply and try different RL learning algorithms. Specifically,  in our
set-up a learning agent (the controller) interacts with an unknown environment
(the opportunistic network). For each content, at fixed time steps, the agent
observes the state of the {\em system} (i.e. the current diffusion of the
content), takes an {\em action} (i.e. decides whether to  inject new content
replicas) and receives a {\em reward}, i.e. a feedback to evaluate the action
taken. Its goal is to learn a {\em policy} (i.e. a mapping from states to
actions) that maximises the long term reward. 

To demonstrate the generality of our approach we also present two instances of
our RL framework by using two well-known and very robust solutions: the  Actor-Critic and the Q-Learning algorithms~\cite{Sutton:1998if,Watkins:1992aa}. These
two approaches represent the trade-off between convergence speed and precision.
In fact, we find that the Actor-Critic learning algorithm, although  slightly
slower in the learning phase, has better performance than the Q-Learning based
approach. Conversely, Q-Learning can be very responsive in the learning phase
but its final accuracy (after the learning phase is over) is worse. 

Through a comprehensive set of simulative experiments we will show that 
solutions based on RL are in general more efficient than other state-of-the-art
approaches. Specifically, with respect to the state-of-the-art benchmarks, our
solution achieves an increased offloading gain (the percentage of offloaded
traffic) of about 20\%.  Moreover, the comparison between the Actor-Critic and
the Q-learning algorithm shows that the former can achieve higher offloading
gain of about 17\%. Finally, our results show that a  solution based on RL  does
not require pre-existing knowledge about the nodes behaviour (e.g., degree on
the contact graph), but is able to automatically learn the most appropriate
control strategy for the offloading task. 

%

This paper extends our previous work in~\cite{Valerio:2014yq}. With respect to~\cite{Valerio:2014yq} we show that the formulation of our offloading procedure
is general enough to be independent from the specific RL algorithm applied. We
also evaluate the performance of two learning approaches (Actor-Critic and
Q-Learning) applied to the offloading problem and, as in our previous work, we
compare their performance to Droid \cite{Rebecchi:2014ai}, a state of  art
solution in cellular offloading. Moreover,  we analyse in more detail, with
respect to~\cite{Valerio:2014yq}, the behaviour of each considered approach in
order to better understand the reasons behind their different
behaviour. 

The rest of this paper is organised as follows. In Section \ref{sec:related} we
review other offloading approaches and we present to what extent they differ
from our offloading solution. In Section \ref{sec:pstatement}   we formalise the
offloading problem as a RL problem. In Section \ref{sec:RL} we provide a brief
background on the RL algorithm we use in this paper  and in Section
\ref{sec:acalg} we show how they are used in the offloading problem.  Section
\ref{sec:perfEval}  presents the performance of all the considered offloading
approaches while Section \ref{sec:conclusions}  concludes the paper.

\section{Related Work}
\label{sec:related}

Over the last years, much effort has been devoted to find good solutions to cope
with the problem of data offloading in cellular networks. In this section we
briefly review the main literature on offloading, and present the main
results regarding data offloading  solutions based on
opportunistic networks for delay tolerant content.  A more complete and
exhaustive survey on state of the art of data offloading techniques in cellular
networks can be found in \cite{Rebecchi:2014ys}.

Offloading traffic from cellular
networks can be done in several ways, depending on the type of constraints  that
should be met for the  content delivery \cite{Rebecchi:2014ys} and the specific
technique applied.
From a technological point of view, a first distinction is between
  \emph{inbound} and \emph{outboud}
  techniques~\cite{Asadi:2014aa}. In the former case,
  offloading
  is performed exploiting the same spectrum of the cellular network,
  while the latter exploits network technologies that use other parts of the
  spectrum, typically in unlicensed bands. Without loss of generality, in the
  rest of the paper we refer to outbound techniques that use WiFi networks to offload
  traffic, although the proposed offloading control mechanism can be used also
  with inbound offloading (most notably, with LTE-D2D, one of the reference
  offloading solutions defined by the LTE standard~\cite{3GPPTSG}).
  
An orthogonal classification is between AP-based and device-to-device
  (D2D) offloading~\cite{Rebecchi:2014ys}. In AP-based offloading (using
  WiFi technologies), end-users located inside a hot-spot coverage area might
  use Wi-Fi connection as an alternative to the cellular network when they need
  to exchange data. In some schemes data transfers are even deferred when no
  WiFi connectivity is available, waiting for the next time the user will move
  close to a WiFi AP~\cite{Mehmeti:2014aa}.  Due to the penetration of WiFi
  Access Points, this is today the most diffused technique, and in its simplest
  case it is already implemented in modern smartphones and tablets, which
automatically switch to a WiFi network when it is
available~\cite{Research:aa}\cite{Mehmeti:2013aa}. In D2D-offloading, cellular
  providers exploit physical co-location and mobility of the users, to send
  content only to a small subset of users via the cellular network, and let
  there users spread the content to other intended destinations through D2D
communications (in the case of WiFi, using the ad-hoc mode).

  Besides, if we take into account the delivery constraints, we can differentiate
between delayed offloading and non-delayed
offloading~\cite{Rebecchi:2014ys}. In the latter case,
additional delays cannot be added to offloaded content (e.g., this is the case
of live streaming video). In the former case, some additional delay is tolerable
by the application (e.g, download of stored content). Clearly, different
  techniques are applied in the two cases (see~\cite{Rebecchi:2014ys} for more
details.)

A first class of offloading techniques based on opportunistic networks
is based on approaches in which  the
dissemination controller has to collect information about nodes' contact rates
in order to come up with the best set of nodes to trigger the dissemination
of contents in the opportunistic network.  
%
%
To the best of our knowledge, Han et al. \cite{Han:2010yu} (then subsequently
extended in \cite{Han:2012qv}) were the first to exploit opportunistic
communications to alleviate data traffic in the cellular network.  In their pioneering work
they propose and  evaluate three algorithms, i.e. \emph{Greedy, Heuristic} and
\emph{Random}, which exploit information about the contact graph of mobile nodes
in order to select the initial set of $k$ mobile nodes in charge to trigger the
opportunistic dissemination of the requested content. 
 %
 Li et al. \cite{Li:2014aa}  propose to treat the selection of seed nodes as a
 maximisation problem in which the utility function to be optimised is subject
 to multiple linear constraints, expressing, for example, the mobility of users,
 traffic heterogeneity and the storage available on mobile nodes. 
 A third solution based on mobility prediction, TOMP (Traffic Offloading with
 Movement Predictions), is proposed by Baier et al.~\cite{Baier:2012ab}. The
 authors try to predict users' mobility in order to  estimate the future
 connectivity opportunities between mobile nodes.  To this end, TOMP  exploits
 information about actual positioning and speed of mobile devices rather than
 connectivity patterns. The framework selects as seed users the nodes that have
 the best future connectivity likelihood with other nodes based on movement
 prediction. TOMP proposes three coverage metrics to predict the future
 movements of nodes: static coverage, free-space coverage, and graph-based
 coverage.

Another class of solutions tackle the problem by exploiting not only the nodes'
mobility characterisation, but also social aspects connected to nodes' mobility.
In the solution  proposed in \cite{Barbera:2014gf}, authors suggest to exploit
social aspects of user mobility to select 
what they call VIP devices that are in charge to spread contents in the
opportunistic network. VIP devices are the most socially important nodes in the
network, i.e. they are able to eventually contact the rest of the network.
Precisely, every time a big amount of data must be transmitted to interested
users, VIP devices act as bridges between the network controller and the
opportunistic network. Authors identify two kinds of VIPs: \emph{global} and
\emph{local}. The former are users globally important in the network while the
latter are important within their social communities. The social importance of
users is estimated through well-known metrics like betweenness
centrality, degree, closeness centrality, and Page Rank.
Following similar principles, \cite{Chuang:2012aa}  takes into account the
social relationships between users of an opportunistic network considering also
information about their social communities in order to select the initial set of
users to bootstrap the spreading of content in the opportunistic
network. Precisely, authors try to minimise both the number of initial seed
users and the latency of the opportunistic dissemination. To this end they
propose community-based opportunistic dissemination, which automatically selects
a sufficient number of initial sources to propagate the content across disjoint
communities in parallel.  In this scheme, mobile end-users are required to
upload periodic information on the most frequent contacts, in order to let the
centralised algorithm choose the best subset of seed users.

All the above approaches are heavily based on the exploitation of social
information collected from the network to select an initial set of seed
users. However, after
the initial
injection of contents, no other actions are taken.

A quite different approach is proposed in \cite{Whitbeck:2012sf}.  Authors
propose Push$\&$Track (PT), an offloading solution based on predefined
performance targets used by the dissemination controller to compute the number
of content replicas to inject.  PT involves also a control loop based on ACK
messages, used to notify the controller every time a content is delivered to a
user.  Precisely, the controller, at fixed time steps, operates content
re-injections in order to force the opportunistic dissemination trend to follow
a target dissemination evolution (i.e. a predefined target percentage of
dissemination at given points in time).  Beyond the number of seed nodes to be
selected over time, the authors investigate methods for node selection like
geographic position or connectivity degree.  Droid (Derivative Re-injection
to Offload Data) \cite{Rebecchi:2014ai}, is the evolution of PT. 
In Droid the authors keep unchanged the ACKs mechanism of PT but substitute the
fixed dissemination strategy with an adaptive mechanism that recomputes at
runtime the offloading targets. 
Precisely, at fixed time steps the controller re-calculates (exploiting the ACKs
that were received) the ideal trend  to reach the full coverage within a maximum time
limit.  Differently from PT, the controller estimates the derivative of the
dissemination evolution over a given short time window in the past, in order to
estimate the trend of progress of dissemination evolution in the future. Based
on this estimate, it checks if content dissemination will reach the total
diffusion within the time limit. Then, it exploits this information to compute
the number of content replicas to be re-transmitted in the network. 
Due to its adaptive and incremental injection strategy, Droid represents a
natural benchmark for our solution, hence, in this paper we compare our results
with Droid. More details on Droid are reported in Section \ref{sec:droid}.

In conclusion, almost all the offloading solutions in the literature consider
significant information about the nodes behaviour (e.g., contact
patterns, social relationships between users) to
select a proper subset of seed nodes to bootstrap opportunistic content
dissemination.
Moreover, except for PT and Droid, just an initial set of nodes is selected with
the consequence of not having any control on the evolution of the dissemination
process.  On the other hand, in solutions like Droid,
decisions of re-injection are taken
using only a very limited part of the information that is available at the
central controller.  In this paper, instead, we argue that the controller can
obtain better results if it learns by experience what is the most appropriate
sequence of re-injection decisions to maximise the coverage while minimising the
cellular network load.  In this way, our solution reaches, autonomously and
independently from the  mobility scenario, higher levels of offloading. 

\section{Problem overview and system assumptions}
\label{sec:pstatement}

We
consider a system composed of $N$ mobile nodes e\-quip\-ped with two network
interfaces: Wi-Fi and cellular (for our purposes, either 3G or LTE would be
equivalent). The former  is devoted to the opportunistic communication
between
devices (and is therefore configured in ad hoc mode) and the latter is used to
communicate with the  dissemination controller. We assume that nodes are
interested in receiving multiple content items over time. Here we consider
contents with  a limited time validity, thus they must be delivered to users within a
deadline. The typical example is automatic download of updated content to users that are subscribed to news websites or
other types of feeds.  We assume a central dissemination controller that decides
whether to send the content items to interested users directly through the
cellular network or indirectly through opportunistic content dissemination.
As in the Push\&Track and Droid systems, we assume that the controller knows the
set of users interested in any given content. The controller has to guarantee
delivery deadlines while minimising traffic over the cellular network.
 To this end, when a new content item needs to be sent to interested users,  the
 controller delegates a set of seeds to
 opportunistically disseminate  the content to all the other peers in the
 network. Without loss of generality we may assume that the opportunistic
 dissemination between nodes is done by means of an epidemic
 algorithm~\cite{Vahdat:2000vf}.  Once a content is delivered to a node, the
 node acknowledges the  controller by sending a message over the
 cellular network, containing its ID and two
 additional pieces of information: the ID of the seed selected by the central controller
 who started the diffusion of that content replica, and the ID of the last
 forwarder-- the reason of sending the two additional IDs will be explained in
 Section \ref{sec:acalg}. Since the ACK message is  lightweight (compared to the
 content size), the impact of this feedback mechanism can be considered
 negligible from the traffic overhead point of view.  After the initial content
 transmission to a limited subset of mobile nodes (seeding),  at fixed time
 steps the controller evaluates the level of dissemination (i.e., the fraction
 of interested users that have received the content item) and decides, based on
 the remaining time until the content item deadline,  whether or not to operate
 a re-injection of the same content in the opportunistic network. Thus, every
 time step the controller calculates how many mobile nodes should be seeded with
 the content.  If the opportunistic dissemination does not reach all interested
 nodes  within the time limit, the controller enters in a ``panic zone'' and
 pushes the content to all the remaining nodes through the cellular network
 (similarly to \cite{Whitbeck:2012sf}).

Note that in our solution the controller implements exactly the same mechanisms
and protocols of Push\&Track and Droid. The key difference is the algorithm
used to decide how many seeds to use at the beginning and during the
dissemination process. To present our solution, we first describe the general
Reinforcement Learning framework on which it is based and the overall logic to
apply it to the offloading problem (Section \ref{sec:RL}). Then, in Section
\ref{sec:acalg} we describe in detail how they are integrated in the mechanisms
explained in this section. 

\section{Reinforcement Learning Algorithms}
\label{sec:RL}

In this section we
introduce the  general principles of Reinforcement Learning  and the two RL
techniques  we used in our work.  We point out that a complete and
formal introduction to Reinforcement Learning is beyond the aim of this paper,
thus we limit our discussion to the most relevant aspects only. A more
complete presentation of this topic can be found in \cite{Sutton:1998if}.

A problem of RL can be modelled as a Markov Decision Process defined by the
tuple $\langle \mc X, \mc A, \phi, \rho \rangle$ where $\mc X$ is the set of
states, $\mc A$ is the set of actions, $\rho: \mc X \times \mc A \rightarrow
\mathbb{R}$ is a reward function and $\phi:\mc X\times \mc A \rightarrow \mc X$
is the state transition function. At each time step $t$, the reward function $\rho$
specifies the instantaneous  reward $r_t$ obtained by taking a given action
$a_{t-1}$ in a given state $x_{t-1}$, while $\phi$ represents the state
transition triggered by the action.  Precisely, the $\phi$ function returns the
state $x_t$ that the system reaches from state $x_{t-1}$ after applying action
$a_{t-1}$.  At each time  step the actions are taken according to another
function called {\em policy} $\pi: \mc X \rightarrow \mc A$.  We anticipate that
in our case the state will include the fraction of interested nodes that have
received the content at time $t-1$, while actions are the number of new seeds that
are created at time $t-1$ (i.e. the number of additional users to which content is
sent directly via the cellular network). Finally, rewards will be the increase
of interested users reached by the dissemination process between time step $t-1$
and time step $t$. In general the function
$\phi$  is 
unknown to the controller because it is impossible to
 model the
real effect of a content injection on the dissemination process in the
opportunistic network without considering any assumptions about the nodes
  encounter patterns (remember that we want to design a scheme that does not
rely on any such assumption).  Thus, the controller just observes a series of state transitions and
rewards. 

The quality of a policy is quantified by the value function $V^{\pi}: \mc X
\rightarrow \mathbb{R}$ defined as the expected discounted cumulative reward
starting in a state $x$ and then following the policy $\pi$:
\begin{equation}
V^{\pi}(x)= E\left[\sum_{t=0}^{\infty} \gamma^t r_{t}\right] \end{equation}
where $\gamma \in [0,1]$ is a discount factor, that reduces the value of the
expected future rewards, to take into consideration the uncertainty in the
evolution of the state in the future.  The goal of the controller is to learn a
policy such that for each state $x$ the expected discounted cumulative reward
is maximised: \begin{equation}
V^{\pi*}(x)= \max_{\pi}E\left[\sum_{t=0}^{\infty} \gamma^t r_{t}\right]
\label{eq:optV} \end{equation}

The solution of Equation (\ref{eq:optV}) can be calculated only if a complete
representation of the environment, over which actions are taken, is known in
advance. 
This might be possible only if mobility patterns over time can be described
through simple closed form expressions, and for simple opportunistic
dissemination schemes whose behaviour can also be described analytically in
simple closed forms.  To cope with more general cases, we use model-free
Reinforcement Learning methods, i.e. RL  approaches that are not based on a
model of the environment.  These methods are known to find very good
approximations of the optimal policy.

In this paper we consider two well known model-free approaches of Reinforcement
Learning: Actor-Critic Learning and Q-Learning. In the following we introduce
both of them, providing details about how they are applied in the context of
cellular data offloading.  \subsection {Actor-Critic Learning}
The Actor-Critic methods are policy-gradient methods based on the simultaneous
on-line estimation of the parameters of two structures: the {\em Actor} and the
{\em Critic}. The Actor corresponds to an {\em action-selection policy} that
maps states to actions in a probabilistic manner, i.e. depending on the current
state, it  tries to select  the best action to take according to the policy. The
Critic is the {\em state-value function} that maps states to expected cumulative
future rewards. Basically, the Critic  is a function predictor that during time
learns the value of a state in terms of the expected cumulative rewards that
will be achieved starting from that state and following the actions taken
(probabilistically) by the Actor. Therefore, the Critic solves a prediction
problem, while the Actor solves a control problem. Although separate problems,
they are solved simultaneously in order to find an approximation of the optimal
policy.  In Actor-Critic methods the Critic evaluates the actions taken by the
Actor and drives the learning process in order to maximise the cumulative
rewards.  In the following we describe both the Critic and the Actor separately.

\subsubsection{Critic}

In our proposed offloading scheme the Critic is designed
by exploiting a {\em temporal-difference} (TD) learning method. TD learning is a
well-known incremental method introduced by Sutton \cite{Sutton:1998if} to
estimate the expected return for a given state. Specifically, when the
transition from the state $x_{t-1}$ to the state $x_t$ and the reward $r_t$ are
obtained, according to the TD theory, the estimation of the value of the state
$x_{t-1}$ is updated in the following way:
\begin{equation} \hat V_{new}^{\pi}(x_{t-1}) = \hat V_{old}^{\pi}(x_{t-1})+
  \alpha\delta_t
\label{eq:updateV} \end{equation}
where $\alpha$ is a  discount parameter. The second part of Equation
(\ref{eq:updateV}) is defined in
Equation (\ref{eq:delta}) and identifies  the utility of action $a_{t-1}$ taken in
state $x_{t-1}$ at time $t-1$. 

\begin{equation}
  \delta_t= r_{t} + \gamma \hat V^{\pi}(x_t) -
  \hat V_{old}^{\pi}(x_{t-1})
  \label{eq:delta}
\end{equation}
with $0<\gamma \leq 1$. The rationale of Equation (\ref{eq:delta})  is to consider
an action very useful if (i) it provided a big reward $r_t$, and if (ii) it
brought the system to a state $x_t$ whose value is much higher than the previous
state $x_{t-1}$. Given this, Equation (\ref{eq:updateV}) is a simple discounted
update of the value of the state $x_{t-1}$, based on the utility of the action
taken.


\subsubsection{Actor}
\label{sec:actor}

Now we describe the policy function through which actions are selected.  The
Actor-Critic mechanisms aims at exploring the state space and avoiding being
trapped in locally optimal actions. Therefore the Actor chooses actions in a
probabilistic (rather than deterministic) way. Precisely, it uses the Gibbs
Softmax method~\cite{Sutton:1998if}, a well known approach in RL for action
selection. According to it, for each state $x_t$, the policy function (i.e., the
probability of selecting action $a_{t-1}$) is given by:

\begin{equation}
  \pi_t(x_{t-1},a_{t-1}) =
  \frac{e^{p(x_{t-1},a_{t-1})}}
  {\sum_{b_{t-1}\in \mathcal A} e^{p(x_{t-1},b_{t-1})}}
  \label{eq:policy}
\end{equation}
%
The probability value of each action is affected by the exponent of the
exponential function, denoted by the function $p(x_{t-1},a_{t-1})$, which is
called \emph{preference function}.
Therefore, learning a policy consists in strengthening or weakening the value of
$p(x_{t-1},a_{t-1})$.  In the Actor-Critic framework, this is done again based
on the utility of having taken action $a_{t-1}$, which is given by the quantity
$\delta_t$ in Equation (\ref{eq:delta}). Specifically: \begin{equation}
p_{new}(x_{t-1},a_{t-1}) = p_{old}(x_{t-1},a_{t-1}) + \beta \delta_{t}
\label{eq:updatePref} \end{equation} where $0<\beta\leq1$. At the beginning,
values of $p$ for all actions are initialised at 0, such that actions are taken
according to a uniform distribution. As we can notice, the preference value is
updated according to the evaluation made by the Critic, or in other words, the
Critic drives the future actions of the Actor.

Summarising, the entire algorithmic procedure for the Actor Critic method is as
follows.  At time $t-1$ the system is in the state $x_{t-1}$. After that, the
Actor draws and executes an action $a_{t-1}$ according to
Equation~(\ref{eq:policy}), and the
new state $x_{t}$ and the reward value $r_{t}$ are observed. The Critic computes
$\delta_{t}$ as in Equation~(\ref{eq:delta}) and updates the estimation of the value
function $\hat V^{\pi}(x_{t-1})$ according to Equation~(\ref{eq:updateV}). Finally, the
policy for state $x_{t-1}$ is updated as in (\ref{eq:updatePref}). This
procedure is  repeated every time step.

\subsection{Q-Learning}

Q-Learning (QL) has been firstly introduced by Watkins
\cite{Watkins:1992aa}. In principle, Q-Learning is similar to Sutton's TD
learning: an agent tries an action at a particular state, and evaluates its
consequences in terms of the immediate reward it receives and its estimate of
the value of the state to which the system is brought.
Differently from the Actor-Critic
method presented before, Q-Learning does not evaluate the value connected to a
state, instead, it evaluates the pair {\em state-action}. In fact Q-Learning
does not estimate the function $V^{\pi}(x)$ like in the Actor-Critic but the
function $Q^{\pi}(x,a)$. Formally, after every state transition, the value for
the state-action pair is updated as follows: \begin{multline} \hat
  Q_{new}^{\pi}(x_{t-1},a_{t-1}) =  \hat Q_{old}^{\pi}(x_{t-1},a_{t-1})  + \\
  \alpha\left[ r_t + \gamma \max_a \hat Q^{\pi}(x_t,a) - \hat
  Q_{old}^{\pi}(x_{t-1},a_{t-1})\right] \label{eq:updateQ} \end{multline} where
$0< \alpha,\gamma<1$ are discount parameters. The intuitive explanation of
Equation (\ref{eq:updateQ}) is that the value of a given pair $(x_{t-1},
a_{t-1})$ is increased based on (i) the obtained reward $r_t$ and (ii) the
difference between the maximum value (over all actions that could be taken
at the next step) for the new state $x_t$ (i.e., $\max_a \hat
Q^{\pi}(x_t,a)$) and the current value, which is an estimate of how much the action has
taken the system to a state with a potentially high value.
The principle behind Q-learning and the Critic part of the Actor-Critic
algorithm is very similar. However, Q-learning is known to converge, in general,
more slowly. Specifically, Q-Learning, under very general assumptions,
asymptotically converges in probability to the optimal action-value function
$Q^*$ independently from the policy $\pi$.  Intuitively, the advantage of AC
over QL is that in the former only one value is computed for each state
(as opposed to each $<$state,action$>$ pair) and this reduces significantly the required
explorations during the learning phase, ultimately achieving quicker convergence
towards the optimal policy.

In this work, Q-Learning algorithm is used in combination with two different
action selection policies: $\varepsilon$-greedy and {\em Softmax}. 

The $\varepsilon$-greedy is one of the simplest action selection policies.
According to $\epsilon$-greedy, when the system is in a state $x$, the
controller selects  with probability $(1-\varepsilon)$ the action with the
maximum accumulated reward for that state and with probability $\varepsilon$ one
of the other actions at random (uniformly), independently of the reward
estimates. 
More formally, let be $\mathcal{U}$ a uniform continuous random variable and
$u_{t-1}$ a sample drawn at time $t-1$.

\begin{equation}
  \pi(x_{t-1}) =
  \left\{\begin{array}{lr} \mathrm{draw\ random\ action } &   \mathrm{if}\,\,
    u_{t-1} <\varepsilon\\
    \underset{a}{\mathrm{argmax}}\ Q(x_{t-1},a) &\mathrm{if}\,\, u_{t-1}\geq \varepsilon
  \end{array}
  \right.
  \label{eq:egreedy}
\end{equation}

The Softmax policy is the same presented for the Actor-Critic approach in
Section \ref{sec:actor} with the only difference that here the exponent of the
exponential function is the $Q$ function. In fact, in the Q-Learning version
each action $a$ is weighted according  to its value for the state $x$ (and not
the overall value of the state as in the Actor-Critic algorithm). Formally, the
probability of the action $a$ for the state $x$ is defined as:

\begin{equation}
  \pi(x_t,a_{t-1}) =
  \frac{e^{Q(x_{t-1},a_{t-1})}}
  {\sum_{\forall a' \in \mc A} e^{Q(x_{t-1},a'_{t-1})}}
  \label{eq:softmaxQ}
\end{equation}

Summarising, the algorithmic procedure for the Q-Learning method is the
following. At time $t-1$ the system is in the state $x_{t-1}$. After having
drawn and executed the action $a_{t-1}$   according to policy (\ref{eq:egreedy})
or (\ref{eq:softmaxQ}), the new state $x_{t}$ and the reward value $r_{t}$ are
observed. Then the action-value for state $x_{t-1}$ are updated
according to Equation~
(\ref{eq:updateQ}). This procedure is repeated every time step. 


\section{Offloading through Reinforcement Learning}
\label{sec:acalg}

In this
section we describe how we exploit the RL framework in the context of data
offloading through opportunistic networking.  In the following we present a
general procedure in which both Actor-Critic and Q-Learning can be plugged in without any
further modification.  In order to apply both RL algorithms  in this context we
need to define: the system state representation, the actions allowed for AC and
QL and the reward function.  In our solution the system state at time $t$ is
represented by the 3-dimensional continuous vector $\boldsymbol
x_t=\{x_{1_t},x_{2_t},x_{3_t}\}\in [0,1]^3$ where  $x_{1_t}$ represents the
dissemination level for the content, $x_{2_t}$ is the fraction of currently used
seeds w.r.t. the total number of  interested nodes (i.e. this quantity is less
than or equal to $x_{1_t}$ depending on the effectiveness of the opportunistic
dissemination) and $x_{3_t}$ denotes the percentage of remaining time for
content delivery before the panic zone.  In our solution the actions represent
the percentage of nodes (still waiting the content)  to be used for the next
re-injection. We have used a discrete set of $11$ actions:
$A=\{a_1, a_2,a_3, \dots, a_{11}\} = \{0,0.01,0.02,\dots,0.1\}$, i.e. we
consider re-injections from $0\%$ to $10\%$ of the (remaining) nodes that should
get the content, in steps of $1\%$. Note that the chosen set of actions is just
one possible setting,  any set of actions is possible, depending on the context.
In fact a cellular operator may configure the system to operate re-injections of
different size, considering context information like the cell's coverage area
and the density of users connected it.  Finally, the reward used to evaluate the action
taken by the RL algorithm is defined in Equation (\ref{eq:reward}).
Specifically:

\begin{equation}
\label{eq:reward}
r_t = - \omega(1-(x_{1_t}-x_{2_{t}})) - (1-\omega)(1-x_{3_t})
\end{equation}
where $\omega$ is a weighting factor ($0 \leq \omega \leq 1$). 
The first term in Equation~(\ref{eq:reward}) measures how much the dissemination process
on the opportunistic network has been effective up to time $t$.  In fact
$1-(x_{1_t}-x_{2_t})$ is the fraction of nodes that have not been reached by the
opportunistic dissemination process. This component, therefore, penalises
actions that lead to using too many seeds (i.e. too low values of $x_{1_t}-x_{2_t}$). The
second component considers the relative time we still have until the content
deadline. It's role is to give more value to a good action if it is taken closer
to the deadline, i.e. when it is needed the most. 
Note that Equation (\ref{eq:reward}) is compliant with the definition of the
reward function in Section \ref{sec:RL}, i.e. it is a function of both the state
where we start from at time $t-1$, and the action taken, as the action leads to
the values of the state at time $t$, which are used to compute the reward. 

As stated in Section \ref{sec:pstatement}, during the validity of a requested
content $c$, our system possibly operates multiple injections of $c$ in the opportunistic
network to speed up the content diffusion.  Now we will describe the  entire
procedure adopted by the controller to drive the opportunistic dissemination
process.  For the sake of presentation clarity, Algorithm \ref{alg:offload}
provides the pseudo-code description of our RL-based offloading scheme.  First,
the system  selects  and transmits the content $c$ to an initial subset of
interested nodes in order to trigger the opportunistic dissemination process.
In order to select the initial set of seeds the RL algorithm draws an action $a$
according to the selected policy $\pi$ (line 11). Precisely, in the AC approach
the Actor, i.e. the action selection policy, draws an action according to
Equation (\ref{eq:policy}), while  QL uses the policy (\ref{eq:egreedy}) or
(\ref{eq:softmaxQ}).  Once the action is selected, our system calculates the
number of new seeds  (line 12), identifies the new seeds (line 13) according to
the result of the function \textsc{GetNewSeeds}(s) (lines 29-32) , and sends the
content to  them (line 14) according to the procedure
\textsc{Offload}($S_{new}$) (lines 33-35).  Finally, the set containing the
active seeds is updated (line 15).  The seeds selection is done considering the
utility of each node. Precisely,  each interested node in the system is
associated to a utility value. Every time a content is delivered to an
interested node, the latter notifies the delivery to the controller together
with the ID of the seed that started the dissemination of that specific copy of
the content and  the ID of the last forwarder. Then, the controller updates the
counters associated to those nodes. In this way, the controller can estimate
over time the utility (identified by the value of the counter associated to a
node)  of each node  in the dissemination process and exploits this information
to identify the most infective nodes to be selected as new seeds (lines 29-32). 
Every time step $t$\footnote{In our system we use discrete time steps, where a
time step is computed as follows $t = s + \tau$, with $0\leq s < t$ and $\tau$
is a fixed quantity (e.g. 5 seconds).}   the system observes the new state
$\boldsymbol x_t$ and the reward value $r_t$ according to Equation
(\ref{eq:reward}) (line 21). It uses this information to  update the estimate of
the value function $\hat V^\pi$ (or $\hat Q^\pi$ in the case of QL) for the
state $\boldsymbol x_{t-1}$ and the policy parameters associated to that state
(line 22) according to procedure \textsc{UpdateRL}($alg$) (44-50). Once the
state-value function  (or the action-value function for the QL algorithm) and
the policy  are updated, we use the latter to draw the next action, i.e. the new
portion of seeds to which content should be sent (line 24). Finally, new seeds
are selected according to the procedure previously described, the content is
offloaded to the new seeds (line 24-26)  and the set of active seeds is updated
(line 27).  When the dissemination process enters the panic zone (line 17) the
controller transmits the content to all the remaining interested nodes.  
\begin{algorithm}[hb!] {\small \caption{ Offloading algorithm.
    \label{alg:offload}} \begin{algorithmic}[1] \State Let $alg$ be  the
      selected RL algorithm (AC or QL) \State Let $c$ be  the requested content
      \State Let $S_t$ be  the set of active seeds, $S_t^c$ the rest of nodes
      waiting for  $c$ and $S_{new}$ the set of new seeds. 
\State Initialise $\boldsymbol x_0 = \{0,0,1\}$ \If{$alg$ is AC} \State
Initialise $ V(\boldsymbol x)=0$, $ p(\boldsymbol x,a)=0\ \forall \boldsymbol
x,a$ \Else \State Initialise $ Q(\boldsymbol x, a)=0\ \forall \boldsymbol x,a$
\EndIf
\State \textbf{For each new content}: \State Draw the action $a_0$ from policy
$\pi$ for the state $\boldsymbol x_0$ \State $s  = a_0 * |S_0^c|$ \State
$S_{new} \leftarrow$ \Call {GetNewSeeds}{$s$} \State \Call {Offload}{$S_{new}$}
\State $S_0 \leftarrow S_0 \cup S_{new}$ \For{each time instant $t$} \If{$t \geq
T$} \State Transmit $c$ to all nodes in $S^c$ \State Stop content dissemination
\EndIf \State Measure $\boldsymbol x_t$, $r_t$ \State \Call{UpdateRL}{$alg$}
\State Draw the action $a_t$ from $\pi_t$ for the state $\boldsymbol x_t$ \State
$s = a_t * |S_t^c|$ \State $S_{new} \leftarrow $ \Call{GetNewSeeds}{$s$} \State
\Call {Offload}{$S_{new}$} \State $S_t \leftarrow S_t \cup S_{new}$ \EndFor
 \Function{GetNewSeeds}{$s$} 
 nodes in $S_t^c$ by their utility value \State return the first $s$ seeds
 \EndFunction \Function{Offload}{$S_{new}$} \State Transmit $c$ to all nodes in
 $S_{new}$ \EndFunction  \Function{UpdateAC}{} \State $\delta \leftarrow r_{t} +
 \gamma V^{\pi}(\boldsymbol x_t) - V^{\pi}(\boldsymbol x_{t-1}) $ \State
 $V^{\pi}(\boldsymbol x_{t-1}) \leftarrow V^{\pi}(\boldsymbol x_{t-1}) +  \alpha
 \delta$ \State $p(\boldsymbol x_{t-1},a_{t-1}) \leftarrow p(\boldsymbol
 x_{t-1},a_{t-1}) + \beta \delta$ \EndFunction \Function{UpdateQL}{} \State
 $Q^{\pi}(\boldsymbol x_{t-1}, a_{t-1}) \leftarrow Q^{\pi}(\boldsymbol x_{t-1},
 a_{t-1}) + \alpha \left[ r_t + \gamma \max_a \hat Q^{\pi}(\boldsymbol x_t,a) -
 \hat Q_{old}^{\pi}(\boldsymbol x_{t-1},a_{t-1})\right]$ \EndFunction
 \Function{UpdateRL}{$alg$} \If {alg is AC} \State UpdateAC() \Else \State
 UpdateQL() \EndIf \EndFunction \end{algorithmic} } \end{algorithm}

\section{Performance Evaluation}
\label{sec:perfEval}

Performance evaluation is accomplished through a set of experiments in different
scenarios. Performance results of our solution are compared with Droid
which is briefly presented in Section~\ref{sec:droid}.  

\subsection{Simulation setup}

In this paper we consider a simulated environment
with $600$ mobile nodes moving in a $3000\times3000 m^2$ area. Each mobile node
represents a mobile device  equipped with two radio interfaces, one simulating
long-range cellular communications (e.g. LTE) and the other one simulating
short-range communications (e.g. WiFi).

The WiFi transmission range is set to 30m. Nodes mobility patterns are generated
according to HCMM \cite{Boldrini10}, a mobility model that integrates temporal,
social and spatial concepts in order to obtain an accurate representation of the movements or real users. Specifically, in HCMM the simulation space is divided in cells
representing different social communities.  Here we consider both a single
community scenario (SC), a two-community scenario (MC2) and a five community
scenario (MC5).  For the single community scenario, all nodes move in the same
cell and have a chance to meet all the other nodes in the cell. In the multi-community scenario we equally split the nodes in two and five separate communities, respectively,
located distant enough to each other such that no border effects may occur, i.e.
nodes moving on the edge of two communities do not fall in the reciprocal
transmission range.  In this setting we consider also the presence of special
nodes, called travellers, that move across the different communities. We use 10
travellers for the MC2 case and 25 travellers for the MC5 case. They represent
the only bridge between nodes of different communities,  if we exclude the
cellular network.  Notice that in the multi-community scenarios, nodes in each
physical community move in a smaller area with respect to the case of single
community, therefore in MC2 and MC5 nodes inside each community have more
contact opportunities  than in SC (although, clearly, the diversity of contacts
for each node is lower). This aspect, as it will be clear in the following, if
carefully exploited by the offloading algorithm, can have a significant impact on the
offloading performance. Finally, all contact traces used in this paper  simulate
one week of nodes' mobility.

\begin{table}[ht!]
  \small
  \caption{Detailed scenario configuration. \label{tab:scenario}}
  \centering
  \begin{tabular}{|c|c|}
    \hline
    \textbf{Parameter} & \textbf{Value}\\
    \hline
    Node speed & Uniform in $[1,1.18m/s]$ \\
    Transmission range & 30m \\
    Simulation area & $3000\times3000 m^2$ \\
    N. of cells & $1\times 1, 3\times 3,6\times 6$\\
    N. of nodes & $600$ \\
    N. of communities & $1,2,5$ \\
    N. of travellers & $0,10,25$\\
    Simulation time & $604800s$\\
    \hline
  \end{tabular}
\end{table}

In our experimental setup content to be delivered to
interested nodes are generated sequentially, i.e.  in our simulations we offload
one content at a time, and
each content must be delivered within a fixed time deadline.  We have
evaluated all the offloading approaches with the following content deadlines:
$1000s$, $300s$, $120s$. During the life time of the content, in order to boost
its dissemination, the controller can operate content re-injections every $5s$
for the $1000s$ and $300s$ content deadlines and every  $2s$  for the $120s$
content deadline. For example, if we consider contents with a delivery
deadline equal to $1000s$, in a week long simulation we measure the offloading performance of 604
sequentially generated contents and for each one the central controller may
operate up to $200$ re-injections.  All results are averaged over $10$ runs on
$10$ different contact traces, all one week long.  
 
\begin{figure*}[ht!]
  \centering{
    \subfloat[1000s, SC, $100\%$ interested nodes]{
      \includegraphics[width=0.33\textwidth]{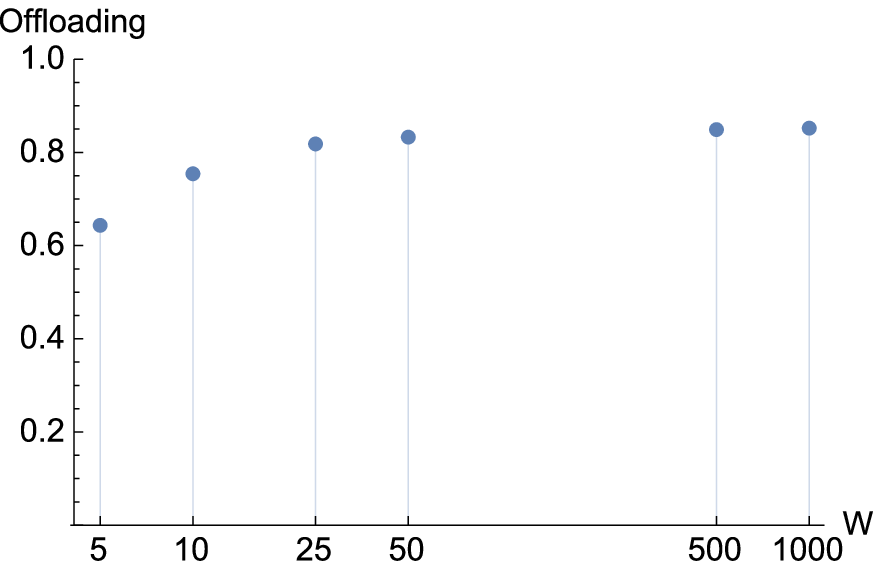}
  }
\subfloat[300s, SC, $100\%$ interested nodes]{
  \includegraphics[width=0.33\textwidth]{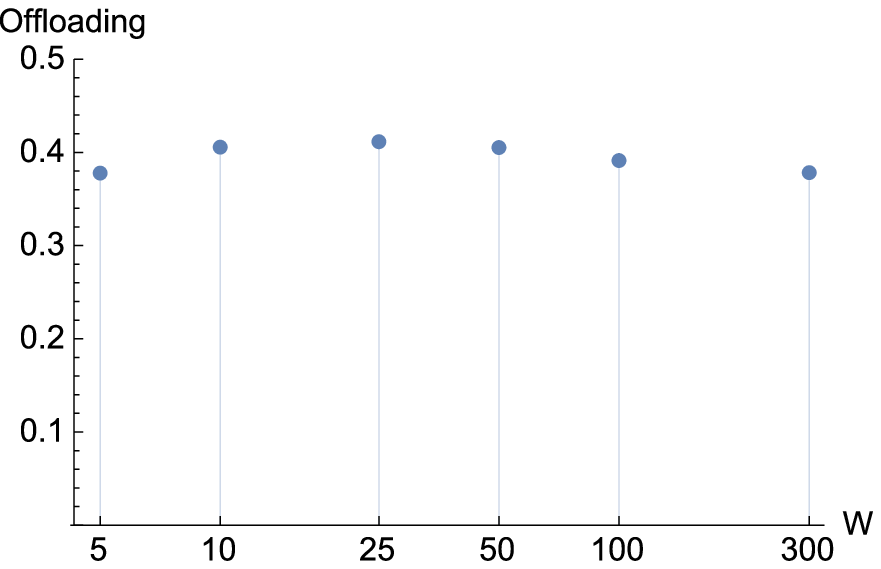} }
  \subfloat[120s, SC, $100\%$ interested nodes]{
    \includegraphics[width=0.33\textwidth]{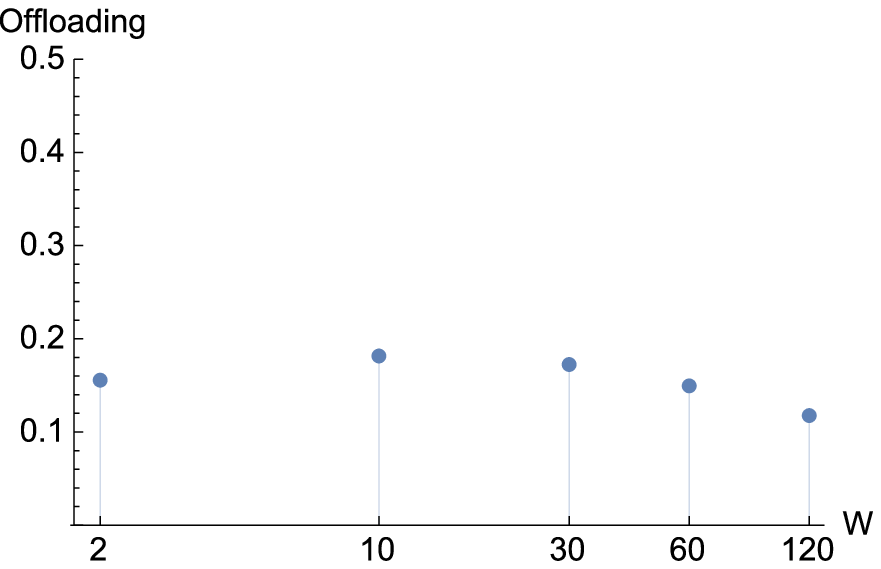}
  }
}
\caption{Droid's sensitivity analysis on the time window W (values on the
  $x$-axis are expressed in seconds) for $c=0.1$. Results for SC where all nodes
  are interested in receiving the content.
\label{fig:DroidSensW}}
\end{figure*}

\subsection{Droid summary}
\label{sec:droid}

For the reader convenience, in this section we briefly describe the Droid's
offloading strategy.  In order to decide if new content replicas should be
injected in the network,  for each time instant, Droid performs the following
actions. First, at time $t$ it computes the slope of the dissemination ratio
($I(t)$) using a discrete derivative estimation method, i.e. as the relative increment of the dissemination ratio in the time window $[t-W,t]$: $$
\Delta_I(t)=\frac{I(t)-I(t-W)}{W}$$ Second, Droid re-injects additional copies
of the content if $\Delta_I(t)$ is below a threshold $\Delta_{lim}$, which is computed on
line as the ratio between the fraction of  nodes waiting the content and the
time remaining ($T$) before the panic zone: $$\Delta_{lim}(t) =
\frac{1-I(t)}{T}$$ The new injection rate $r_{inj}(t)$ is computed as:

\begin{displaymath}
  r_{inj}(t) =
  \left\{
    \begin{array}{lr}
      c, & \Delta_I(t) \leq 0 \\
      c (1-\frac{\Delta_i(t)}{\Delta_{lim}(t)}), & 0<\Delta_i(t)\leq \Delta_{lim}(t)\\
      0, & \Delta_I(t) > \Delta_{lim}(t).
    \end{array}
  \right.
\end{displaymath}
where $0\leq c \leq 1$ is a clipping
value used to limit the amount of content replicas to inject.  Third,  the
number of new seeds $R(t)$ is computed as $$R(t) = \lceil (1-I(t)) \times N(t)
\times r_{inj}(t)\rceil$$ where $N(t)$ is the number of nodes interested in
the content.  Finally, Droid selects  $R(t)$  new nodes using a uniform
distribution and offloads the content replicas. 

In order to understand the results in Section~\ref{sec:expRes},  it is important
to point out that the performance of Droid is affected by the value of
its parameters. As we can see from the above description, its behaviour is
controlled by the clipping value $c$ and the width of the time window $W$.
Between the two, we noticed in experiments that keeping the clipping value
fixed, Droid performance is very sensitive to the value of $W$. In Figure
\ref{fig:DroidSensW}, we reported the mean offloading performance obtained in
the SC scenario for all the content deadlines. On the $x$-axis we have the time
window $W$ and on the $y$-axis the achieved offloading ratio (see
Equation
(\ref{eq:offRatio}) in Section \ref{sec:expRes}). Interestingly, in every
considered scenario we notice that there exists a different value of $W$
corresponding to the higher offloading result, and this optimal value changes in
different scenarios. This is just an example of a more general behaviour we have
observed, i.e. the fact that the parameter $W$ (for any given $c$) needs to be
tuned based on the specific scenario where Droid is used. One of the advantages
of using a RL-based scheme is that such tuning is not needed. 
\begin{figure*}[ht!]
  \subfloat[1000s, SC, $100\%$ interested nodes]{
    \includegraphics[width=.33\textwidth]{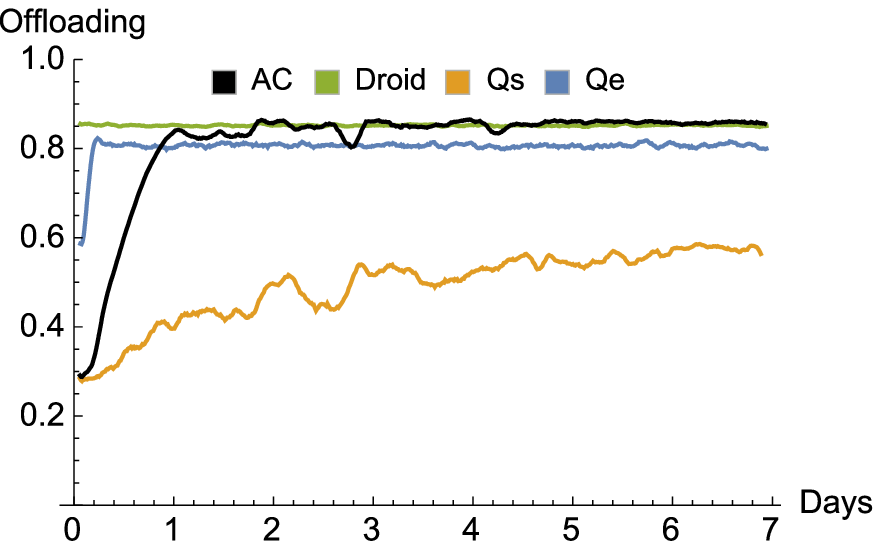}
    \label{fig:AllSC1000_100} }
  \subfloat[300s, SC, $100\%$ interested nodes]{
    \includegraphics[width=.33\textwidth]{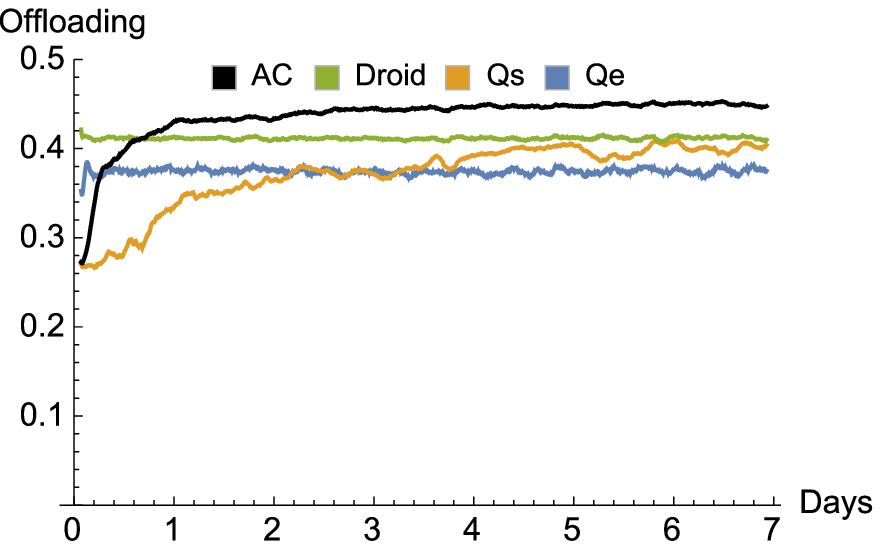}
    \label{fig:AllSC300_100} }
  \subfloat[120s, SC, $100\%$ interested nodes]{
    \includegraphics[width=.33\textwidth]{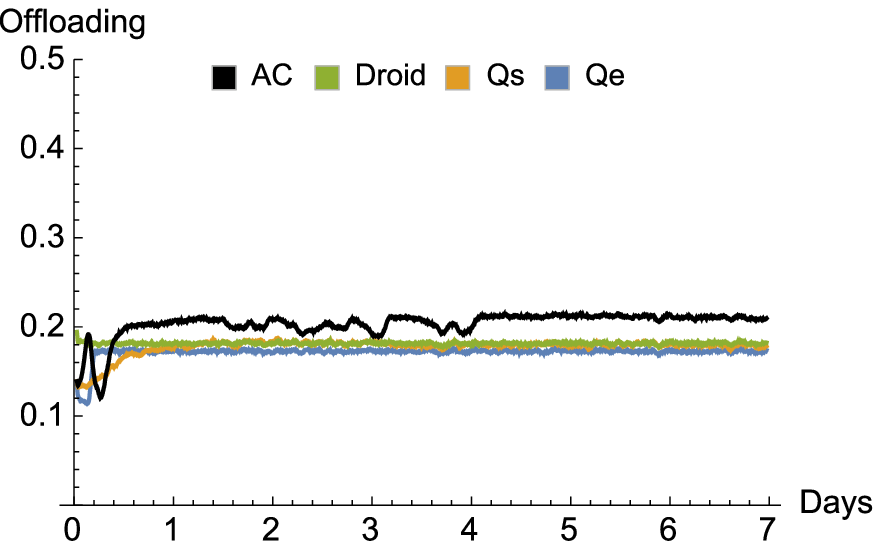}
    \label{fig:AllSC120_100} }
  \caption{Performance evaluation in single community
    with $100\%$ of interested nodes. Considered content deadlines are $1000s$
    $300s$ $120s$. \label{fig:AllSC100} }
\end{figure*}

\subsection{Offloading Efficiency}
\label{sec:expRes}

For the sake of simplicity
from now on we will use  the acronyms AC, Qe, Qs and DR  for Actor-Critic,
Q-Learning with $\varepsilon$-greedy, Q-Learning with Softmax  and Droid,
respectively.  In light of what pointed out in Section \ref{sec:droid},
regarding Droid we report only its best performance for each scenario.  To
evaluate the performance of our approach we compare the evolution of the
\emph{offloading ratio} obtained by all the considered offloading schemes. 
The offloading ratio is defined as percentage of nodes that received the content
from the opportunistic network only. More formally, let denote with $I_c$ the
number of nodes interested in receiving the content $c$. Let $S_c$ be the total
number of seeds used by the central controller for the content $c$ and $P_c$ the
number of interested nodes that received the content $c$ through a unicast
transmission in the \emph{panic zone}. The offloading ratio is computed as
follows:

\begin{equation}
  \mathrm{Offloading\ ratio} = 1 - \frac{S_c + P_c}{I_c}
  \label{eq:offRatio}
\end{equation}
%
 This is a general performance figure, that does not measure the exact amount of
 offloaded traffic in terms of bytes. We prefer anyway to use the offloading
 ratio as the main performance index, because the actually offloaded traffic
 would depend on the specific technologies used for the cellular and
 opportunistic network, the amount of resources allocated by mobile nodes to
 support the offloading process, the size of the content to be delivered. Using
 the offloading ratio abstracts all these dependencies, and gives anyway a
 good indication of the possible actual offloading irrespective of the specifics
 of the network technologies and the application-level traffic.  Furthermore, in
 order to  understand the internal behaviour of the considered offloading
 methods in different scenarios, we also analyses what actions are taken by the
 different algorithms, over time. Specifically, we analyse the evolution over
 time of the frequency with which any specific action is taken.

\subsubsection{Single Community Offloading performance}
\label{sec:SCperf}

We start by considering the SC scenario, assuming that all nodes are interested
in the content items. Figure \ref{fig:AllSC100} presents the results of the
different offloading algorithms. A first general observation is that in absolute
terms the offloading performance of all the approaches is strongly affected by 
the deadline for the delivery of a content. In fact, for a given mobility
pattern,  the more we reduce the content deadline the more the absolute
offloading ratio of all the approaches decreases. This is intuitive, as
for shorter deadlines the time for opportunistic diffusion of contents is lower,
and therefore more nodes need to be served in the panic zone. This also
suggests that for any
scenario there is a limit for the amount of traffic that can be offloaded from
the cellular network, and the objective of all the considered approaches is to
regulate the controller behaviour in order to approximate as much as possible
that limit.

Let us now analyse the obtained results  for different content deadlines. As we
can see from Figure \ref{fig:AllSC1000_100}, with a quite long deadline for
content delivery ($1000s$) after almost 1 day spent for  learning the best
injection policy, AC is able to offload as much traffic as the best fine tuned
configuration of DR. Qe,  thanks to the $\varepsilon$-greedy learning algorithm
shows a steeper learning phase w.r.t AC -- the learning phase reaches stability
just after few hours. However,  following a greedy policy does not pay in the
long term because after some time the learning phase process blocks into a
solution that is not optimal, and this results in the fact that Qe achieves only
$80\%$ offloading ratio, while other non-greedy policies are able to offload more.
Qs , thanks to the Softmax action selection policy, does not get blocked
in some local optimum. However its learning phase is so slow that it is not able
to find a stable solution within the simulation time, leading to very poor
performance. 

For  shorter deadlines (see Figure
\ref{fig:AllSC300_100}-\ref{fig:AllSC120_100}), we can notice two interesting
facts. First, AC proves to be able to well adapt its re-injection strategy and
outperforms all the other solutions. Second, the performance gain of DR w.r.t Qe
and Qs gets smaller and smaller as long as the content deadline gets
shorter. For a content deadline equal to $300s$ we notice an inversion of
performance between Qe and Qs, and Qs is able to eventually outperform Qe. 
Finally, for very short content deadlines ($120s$) we find no more differences
between DR, Qe and Qs while AC shows to be more adaptable to varying content
deadlines, and outperforms all of them.
Average offloading ratios are reported in Table \ref{tab:SC100}.

\begin{figure*}[ht!]
  \subfloat[1000s, SC, $20\%$ interested nodes]{
    \includegraphics[width=.33\textwidth]{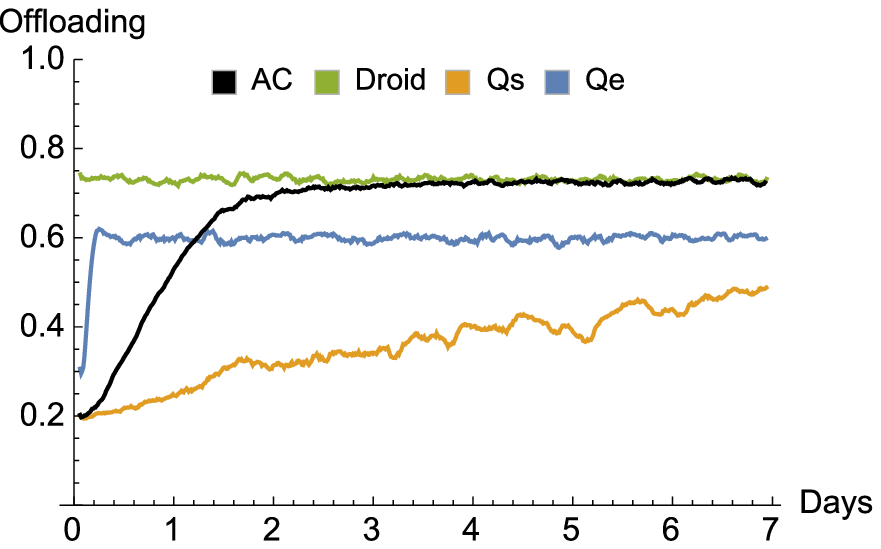}
    \label{fig:AllSC20_1000} }
  \subfloat[300s, SC, $20\%$ interested nodes]{
    \includegraphics[width=.33\textwidth]{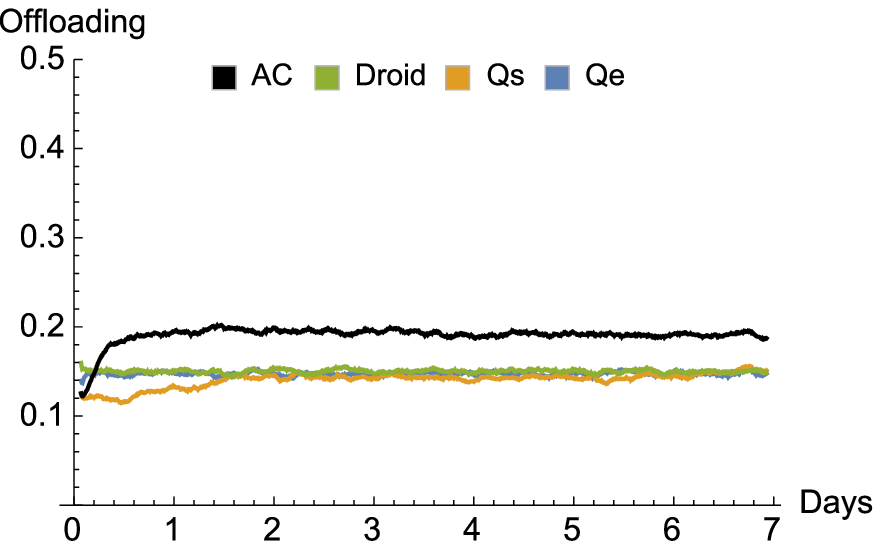}
    \label{fig:AllSC20_300} }
  \subfloat[120s, SC, $20\%$ interested nodes]{
    \includegraphics[width=.33\textwidth]{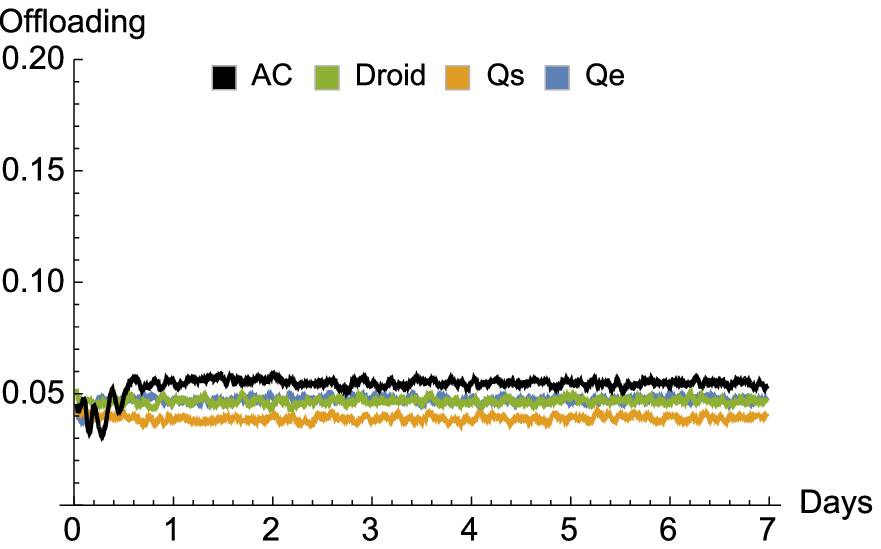}
    \label{fig:AllSC20_120} }
  \caption{Offloading curves in SC scenario with
    $20\%$ of interested nodes\label{fig:AllSC20}}
\end{figure*}

These results already highlight a general feature of the different considered
approaches. Specifically, in the case of long deadlines, there are fewer content
items generated. This results in the fact that Qs has not enough information to
sufficiently discriminate between the possible $<$state,action$>$ pairs, and
therefore takes sub-optimal actions. When deadlines get shorter (and therefore
more content items are handled) this effect becomes less and less impactful, and
therefore Qs achieves better and better performance, eventually outperforming
Qe. As anticipated, the strong point of Qe is that it is quick to find
``relatively good'' actions, although it gets easily trapped in local optima. AC
shows to take the best of both QL algorithms. By using a soft-max policy to
explore possible actions (which is the same policy used by Qs), it avoids being
trapped in local optima as Qe. However, thanks to the different value function,
which gives value only to the state reached, and not the pair
$<$state,action$>$, it is able to learn much quicker which actions to take, similar
to Qe. Finally, with respect to DR, learning algorithms outperform it as soon as
a sufficient number of content items is used for learning (i.e., for shorter
deadline). However, for longer deadlines AC is able to match the performance of
the optimal DR configuration, without requiring any tuning.
\begin{table}
  \caption{ Offloading ratio for SC, $100\%$ of interested nodes.
    \label{tab:SC100}}
  \begin{center}
    \begin{tabular}{|c|c|c|c|c|}
      \hline
      Delivery deadline & AC & DR & Qe & Qs  \\
      \hline 1000s & $85\%$ & $85\%$ & $80\%$ & $53\%$  \\
      300s &  $44\%$ & $41\%$ & $37\%$ & $39\%$\\
      120s & $20\%$ & $17\%$ & $17\%$ & $18\%$\\
      \hline
    \end{tabular}
  \end{center}
\end{table}

We tested the performance of the compared solutions also in a case
  where the number of interested users is reduced, i.e. it is only $20\%$ of the
  simulated users.
We point out that, in this scenario, only nodes that are interested in the
content are eligible for being seed. Moreover, in this experiment, we allowed
the other (non-interested) nodes in the community to collaborate to the
opportunistic dissemination process.  This experimental setup represents a more
challenging scenario because of the different number of possible seeds. Before
comparing the ``100\% case" with the ``20\% case" in detail, notice that the
comparison between the different algorithms presented for the 100\% case holds
also here. For long deadlines AC matches DR and both outperform QL. As deadlines
get shorter and shorter (i) Qs catches up over Qe; (ii) eventually Qs matches
DR and (iii) AC becomes the most efficient solution.

Let us now compare the 100\% case and the 20\% case more in detail.
First of all, let us assume to use an equal number of seeds. As all nodes
participate in the dissemination process, the opportunistic dissemination
process would be (stochastically) equivalent in the two cases. This means that,
the 20\% interested nodes in the second scenario would have the same probability
of receiving the content via the opportunistic network before the deadline in
the two cases. Therefore, the average number of nodes in the ``20\% group" that
will receive content via offloading would be the same. However, to achieve this
we would need to use the same number of seeds in the two cases, and this means
that the offloading ratio would be in general \emph{lower} when fewer nodes are
interested in the content. On the other hand, using fewer seeds in the 20\% case
might not help, because the dissemination process would be slower, and more
interested nodes might enter the panic zone without having received the content
yet. This general behaviour is evident for short content deadlines (compare
Figure \ref{fig:AllSC20_300} and \ref{fig:AllSC20_120} with  Figure
\ref{fig:AllSC300_100} and \ref{fig:AllSC120_100}). On the other hand, when
content deadlines are longer (compare Figure \ref{fig:AllSC20_1000} with
\ref{fig:AllSC1000_100}) then even using a lower number of seeds would be
sufficient to sustain the same offloading ratio achieved in the 100\% case.
Average offloading ratios are reported in Table \ref{tab:SC20}.

\begin{table}[h!]
  \caption{ Offloading ratio for SC, $20\%$ of interested nodes.
    \label{tab:SC20}}
  \begin{center}
    \begin{tabular}{|c|c|c|c|c|}
      \hline
      Delivery deadline & AC & DR & Qe & Qs  \\
      \hline
      1000s & $72\%$ & $72\%$ & $59\%$ & $40\%$  \\
      300s &  $19\%$ & $14\%$ & $14\%$ & $14\%$\\
      120s & $6\%$ & $4\%$ & $4\%$ & $3\%$\\
      \hline
    \end{tabular}
  \end{center}
\end{table}

In order to push all these approaches to their limit we performed an even more
challenging experiment in the SC scenario. As in the previous experiment, the
number of nodes interested in receiving the content is $20\%$ of the total
number of nodes, but now all the other $80\%$ of nodes do not contribute to the
opportunistic dissemination process. Specifically, in this setting the central
controller has to cope with the situation in which the effect of a content
(re-)injection may arrive after a quite long delay because much fewer nodes
participate to the opportunistic dissemination process. In this case for the
offloading controller it is even more challenging to find the right compromise
between the number of seeds, the resulting dissemination process and the time
allowed before the deadline. As we can see from Table \ref{tab:SC20nc}, any
algorithm is able to offload some content only for very long deadlines (i.e.,
when the opportunistic dissemination has sufficient time to reach interested
nodes). As in the former experiments, AC outperforms the other learning
approaches and also Droid.

\begin{table}
  \caption{ Offloading ratio for SC, $20\%$ of interested nodes. Non collaborative
    dissemination.\label{tab:SC20nc}}
  \begin{center}
    \begin{tabular}{|c|c|c|c|c|}
      \hline
      Delivery deadline & AC & DR & Qe & Qs \\
      \hline
      1000s & $33\%$ & $26\%$ & $18\%$ & $23\%$ \\
      300s &  $-$ & $-$ & $-$ & $-$\\
      120s &  $-$ & $-$ & $-$ & $-$\\
      \hline
    \end{tabular}
  \end{center}
\end{table}

  \subsubsection{Multi-community offloading performance}
\label{sec:multicommperf}

\begin{figure*}[ht!]
  \subfloat[1000s, MC, $100\%$ interested nodes]{
    \includegraphics[width=.33\textwidth]{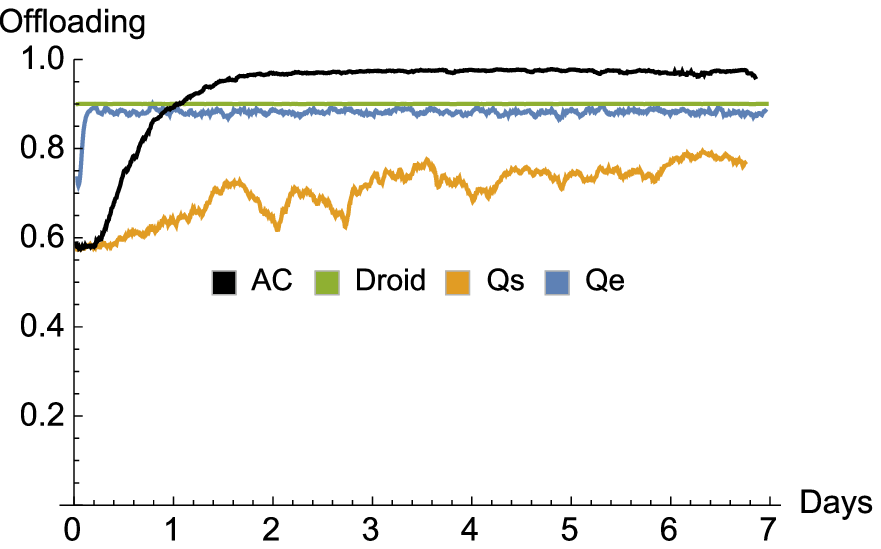}
    \label{fig:MC21000} }
  \subfloat[300s, MC, $100\%$ interested nodes]{
    \includegraphics[width=.33\textwidth]{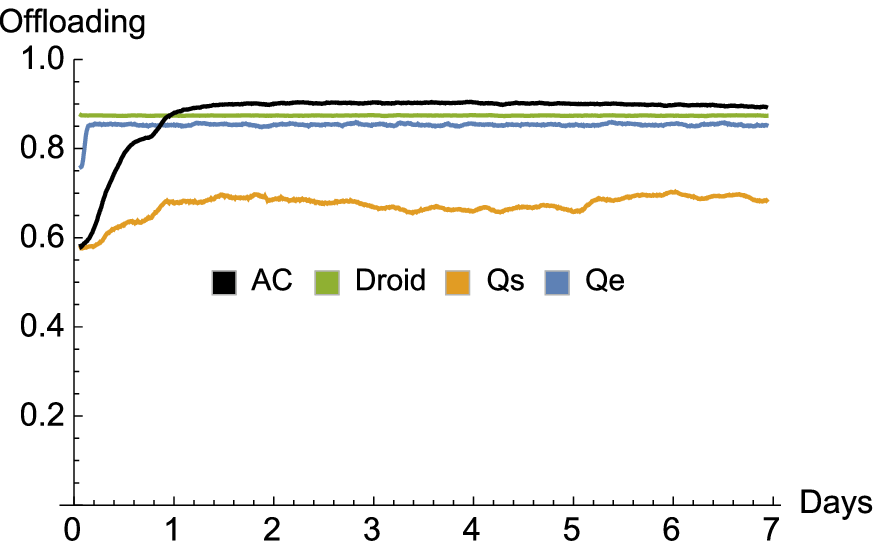}}
  \subfloat[120s, MC, $100\%$ interested nodes]{
    \includegraphics[width=.33\textwidth]{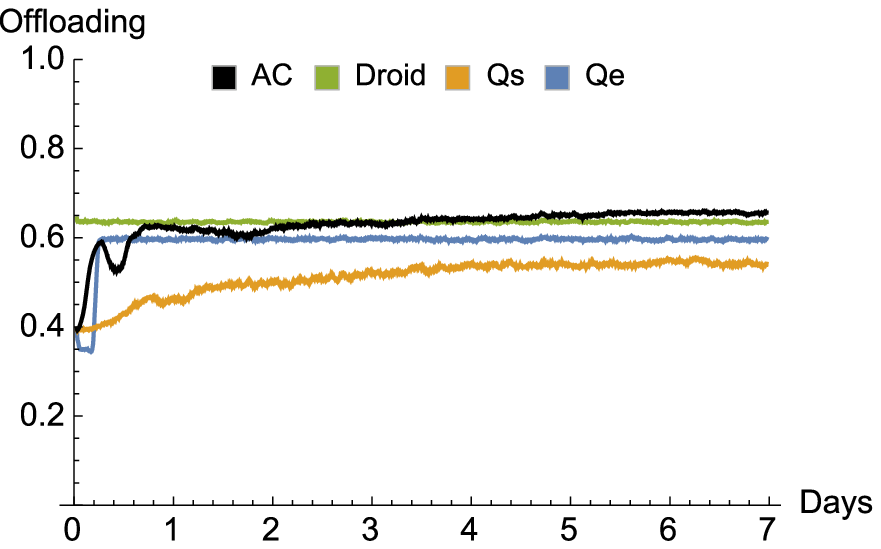}}
  \caption{Offloading curves in MC2 scenario, $100\%$ interested
  nodes\label{fig:MC2}}
\end{figure*}

Now we show the performance of the offloading algorithms in two multi community
scenarios.  The motivation for this kind of experiments is to test the efficiency of the different approaches stress with heterogenous mobility settings. In fact, though we keep fixed the
simulation area and mobility parameters, splitting the nodes in $2$ and $5$
physical groups has the effect of increasing the contact rate between nodes
because the area in which each community moves is smaller.  Thus, we expect that in these
cases more traffic can be offloaded.  As expected, looking at Table
\ref{tab:MC2}, the overall offloading ratios are higher than those obtained in
the SC scenario. Regarding the performance of each approach we notice that also
in this case AC  shows the best adaptability to this quite fast dissemination
process. Looking at Figure \ref{fig:MC2} we notice several similarities with
respect to the SC case. AC typically outperforms  the other approaches while
both versions of Q-Learning are not able to offload more than DR. A notable
difference is that in this case Qe still outperforms Qs when the content
deadline is 300s and 120s. This is due to the fact that the increased contact
rate boosts the performance of Qe.
In the MC5 scenario, the contact rate is even higher that in MC2, therefore all
the approaches can significantly reduce  their
(re-)injections. As reported in Table \ref{tab:MC5}, and similarly to the previous scenarios, AC applies the best re-injection policy, offloading up to $18\%$ more traffic than
its competitors.

\begin{table}[th!]
  \caption{ Offloading ratio for MC2, $100\%$ of interested nodes. \label{tab:MC2}}
  \begin{center}
      \begin{tabular}{|c|c|c|c|c|}
	\hline
	Delivery deadline & AC & DR & Qe & Qs\\
	\hline
	1000s & $97\%$ & $90\%$ & $88\%$ & $75\%$  \\
	300s &  $90\%$ & $87\%$ & $85\%$ & $69\%$\\
	120s &  $65\%$ & $64\%$ & $60\%$ & $54\%$\\
	\hline
      \end{tabular}
    \end{center}
\end{table}

\begin{table}[ht!]
  \caption{ Offloading ratio for MC5, $100\%$ of interested nodes. \label{tab:MC5}}
  \begin{center}
    \begin{tabular}{|c|c|c|c|c|}
      \hline
      Delivery deadline & AC & DR & Qe & Qs  \\
      \hline 1000s & $95\%$ & $90\%$ & $90\%$ & $77\%$\\
      300s &  $95\%$ & $89\%$ & $89\%$ & $77\%$\\
      120s &  $82\%$ & $81\%$ & $77\%$ & $67\%$\\
      \hline
    \end{tabular}
  \end{center}
\end{table}

\subsubsection{Dynamic scenario performance}

Finally, we test both AC and DR in
a more dynamic scenario. We want to simulate a situation where
both nodes' mobility and deadlines for the contents delivery change at a certain
point in time. The aim of this experiment is to evaluate how AC and Droid adapt
their behaviour when the mobility and network scenarios change. It is important to point out that the initial parameter setting in Droid is kept fixed during the experiment and it is not re-tuned after the mobility scenario changes. As explained in the following, it is infeasible for the central controller to have a static mapping between the best fine tuned configuration of Droid and all possible mobility scenarios. We ran AC and DR in the multi-community scenario MC5
with a content delivery deadline equal to $1000s$, then we changed the underlying
mobility scenario from MC5 to SC and the content delivery deadline to $300s$,
keeping unchanged the internal state of the AC and DR algorithms. 

\begin{figure}[h!]
  \centering
  \subfloat[MC5 (1000s) - SC (300s), 100\% interested users]{
    \includegraphics[width=.4\textwidth]{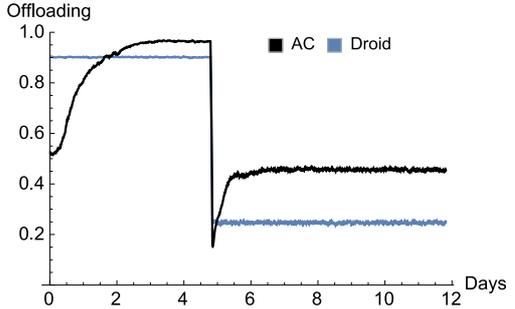}
    \label{fig:dyn100} }\\
  \subfloat[MC5 (1000s) - SC (300s), 30\% interested users]{
    \includegraphics[width=.4\textwidth]{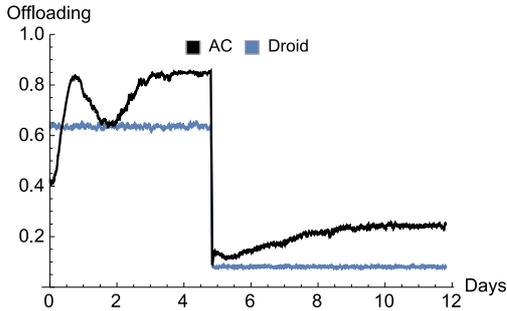}
    \label{fig:dyn30} }
  \caption{Offloading curves in a dynamic scenario. Nodes'
  mobility is MC5 for the first 5 days and SC for the rest. Content deadlines
  are 1000s for MC5 and 300s for SC. Interested users are 100\% (a) and 30\% (b)
  \label{fig:dyn1000}}
\end{figure}

In Figure~\ref{fig:dyn100} we show the performance of AC and DR in a scenario
where 100\% of mobile users are interested in receiving the content. As we can
see, the curves of both approaches in the first part of the plot (the first 5
days) show the same offloading ratio reported in Table~\ref{tab:MC5}. This is
due to the fact that we set the same configuration used in the MC5 experiment
for both  Droid and AC (AC  and DR offload up to 97\% and 90\%, respectively).
After the change of mobility scenario and content deadline we see that
for AC the
learning phase restarts and achieves the same performance obtained in the SC
scenario presented in Section~\ref{sec:SCperf}, i.e. $44\%$. Conversely,  Droid
is able to offload only $24\%$ (17\% less traffic offloaded with respect to
results in Table~\ref{tab:SC100}). This  result shows that the performance of
Droid is strongly connected to a fine-tuning of its
parameters based on the specific scenarios.
This means that if we consider situations where the nodes
mobility and content delivery deadlines might dynamically change, reconfiguring
its parameters for every scenario may become unsustainable. Offloading ratios are
reported in Table~\ref{tab:MC5SC100}.

We performed  another experiment in which not only the mobility scenario changes
(as in the previous experiment first MC5 followed by SC with 1000s and 300s of content delivery
deadlines, respectively), but we also reduce the number of interested users to 30\% of
the total number of users. The results  shown in Figure~\ref{fig:dyn30} confirm
what already presented in Figure~\ref{fig:dyn100}. In both mobility scenarios AC
proves superior adaptivity and performance,  offloading 20\% and 16\% more
traffic than DR in MC5 and SC, respectively. Offloading ratios are reported in
Table~\ref{tab:MC5SC30}.

\begin{table}[ht!]
  \caption{ Offloading ratio for MC5 followed by SC, $100\%$ of
    interested nodes. \label{tab:MC5SC100}}
    \begin{center}
      \begin{tabular}{|c|c|c|}
	\hline
	Delivery deadline & AC & DR  \\
	\hline
	MC5, 1000s & $95\%$ & $90\%$ \\
	SC, 300s &  $44\%$ & $24\%$\\
	\hline
      \end{tabular}
    \end{center}
\end{table}

\begin{table}[ht!]
  \caption{ Offloading ratio for MC5 followed by SC, $30\%$ of interested nodes.
      \label{tab:MC5SC30}}
  \begin{center}
    \begin{tabular}{|c|c|c|}
      \hline
      Delivery deadline & AC & DR  \\
      \hline MC5, 1000s & $84\%$ & $63\%$ \\
      SC, 300s &  $24\%$ & $8\%$\\
      \hline
    \end{tabular}
  \end{center}
\end{table}

\subsubsection{Discussion}

Summarising, in light of the presented experimental
results, we come up with the following key findings:
\begin{itemize}
  \item in general, a RL algorithm (and AC in particular)
      guarantees higher offloading with respect to state-of-the art
      approaches i.e., Droid.
      Moreover, AC is very adaptive and provides the best offloading ratio
      consistently across all considered scenarios. This is important because AC
      does not need fine tuning, and is therefore a very flexible offloading
      approach.
  \item Offloading efficiency is directly correlated with the
      amount of interested users, even in fully cooperative cases, where also
    non-interested nodes contribute to disseminate content.
  \item Among the
      tested RL algorithms, AC is the best solution because it is almost as
      quick as Qe in learning, and as good as Qs in avoiding local minima.
  \item Mobility patterns with higher contact rates improve the offloading
	performance across all tested algorithms, and reduce the advantage of Qs
	compared to Qe in case of short deadlines.
\end{itemize}

\begin{figure*}[ht!]
  \centering
  \subfloat[AC $85\%$ Offloading]{
    \includegraphics[width=0.35\textwidth]{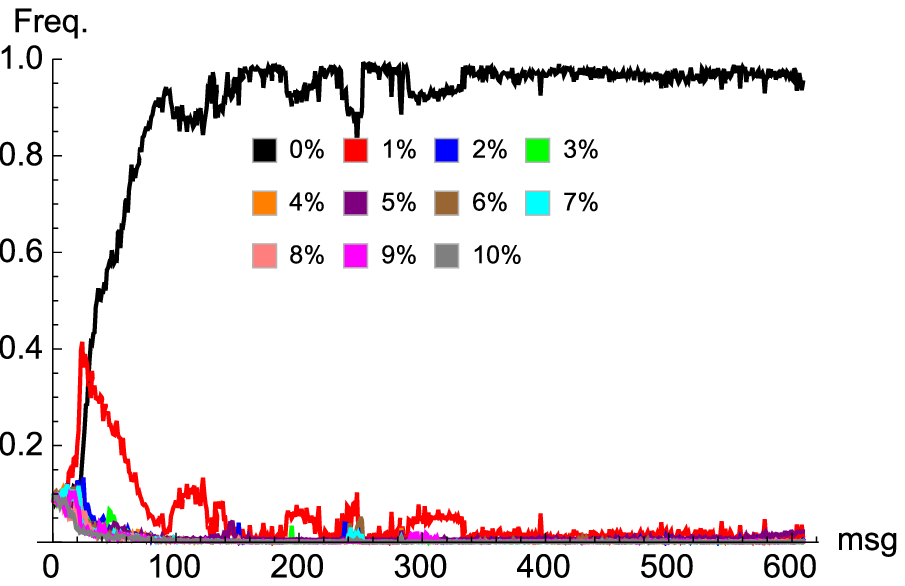}
  }
  \subfloat[DR $85\%$ Offloading]{
    \includegraphics[width=.35\textwidth]{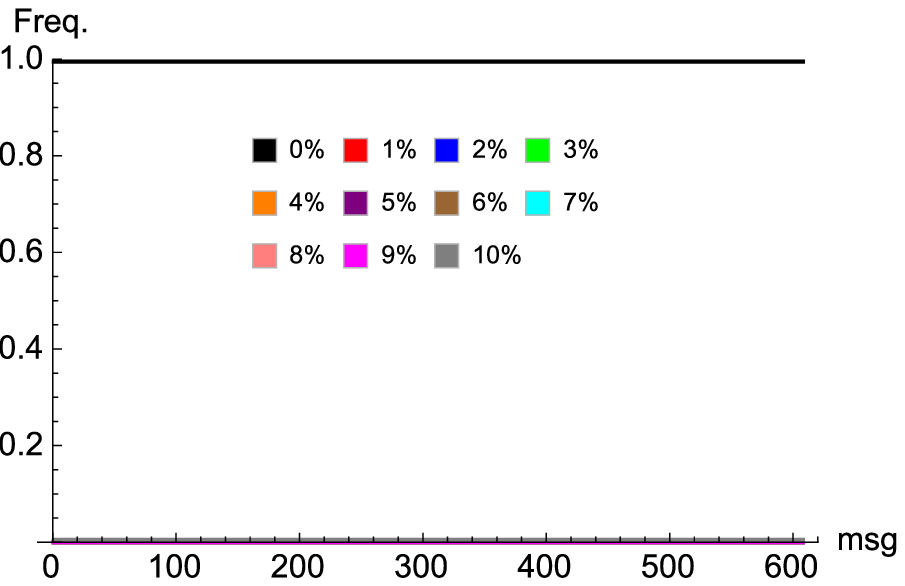}
  }\\
  \subfloat[Qe $80\%$ Offloading]{
    \includegraphics[width=0.35\textwidth]{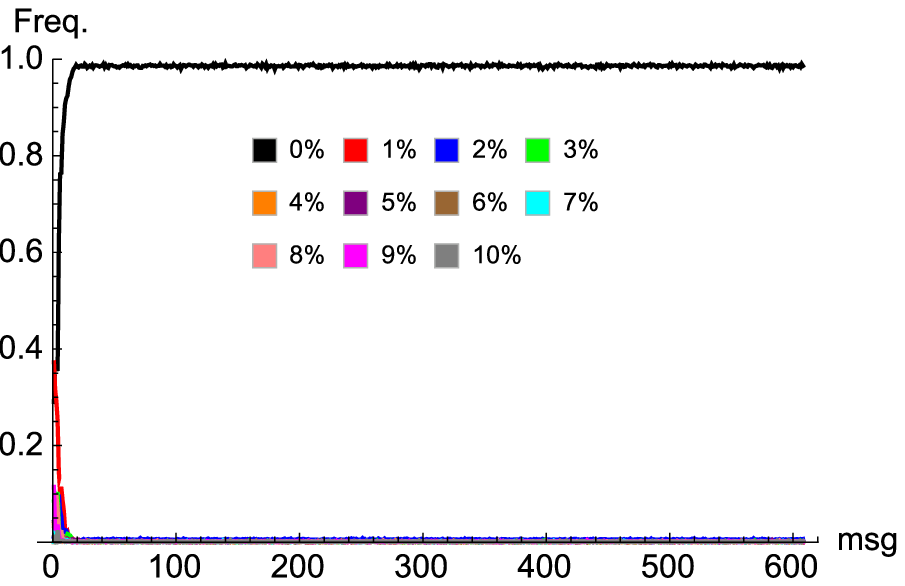}
  }
  \subfloat[Qs $53\%$ Offloading]{
    \includegraphics[width=0.35\textwidth]{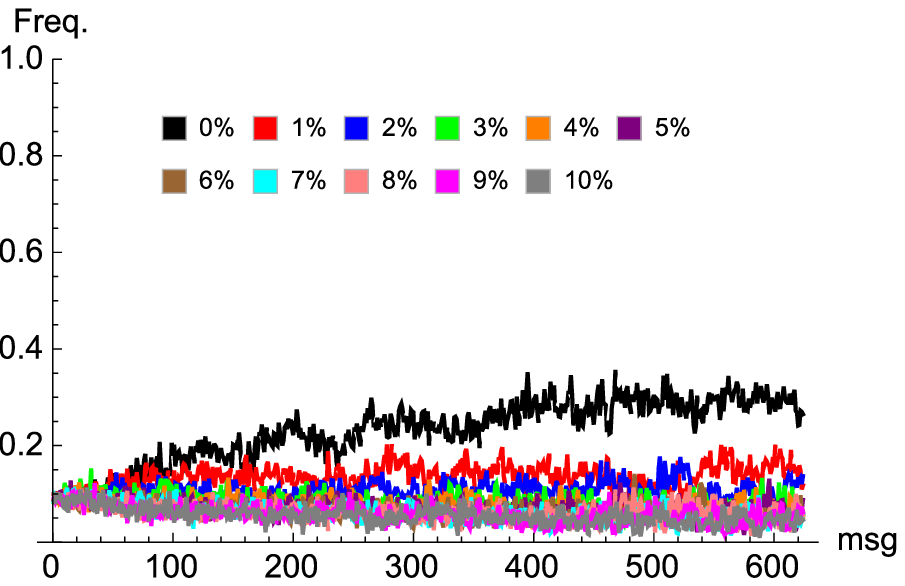}
  }
  \caption{Actions frequencies curves. Deadline 1000s, SC, $100\%$ interested
    nodes\label{fig:actions1000}}
\end{figure*}

\begin{figure*}[ht!]
  \centering
  \subfloat[AC $44\%$ Offloading]{
    \includegraphics[width=0.35\textwidth]{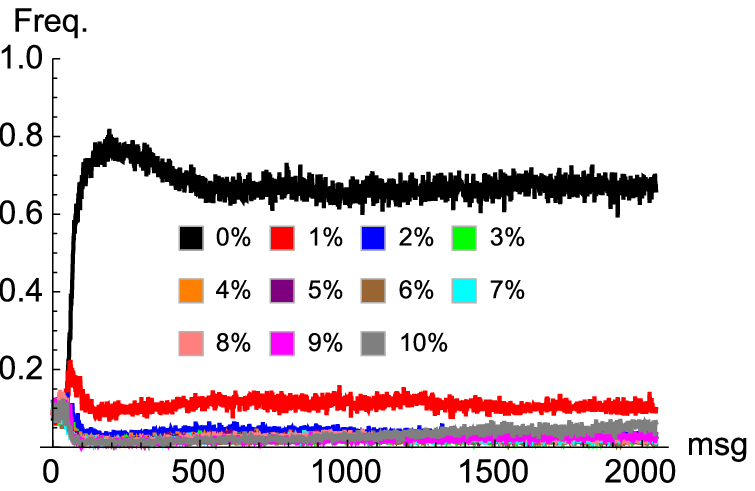}
  }
  \subfloat[DR $41\%$ Offloading]{
    \includegraphics[width=0.35\textwidth]{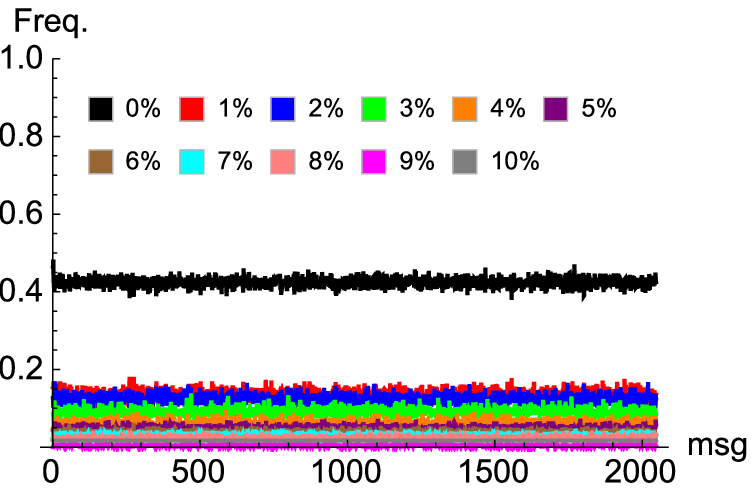}
  }\\
  \subfloat[Qe $37\%$ Offloading]{
    \includegraphics[width=0.35\textwidth]{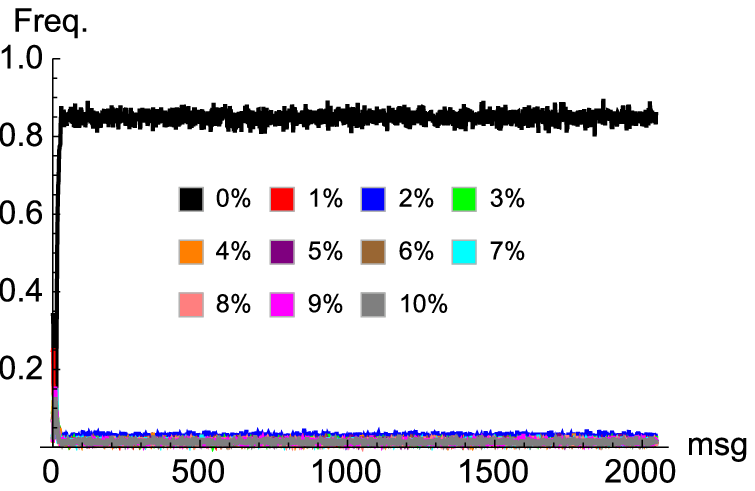}
  }
  \subfloat[Qs $40\%$ Offloading]{
    \includegraphics[width=0.35\textwidth]{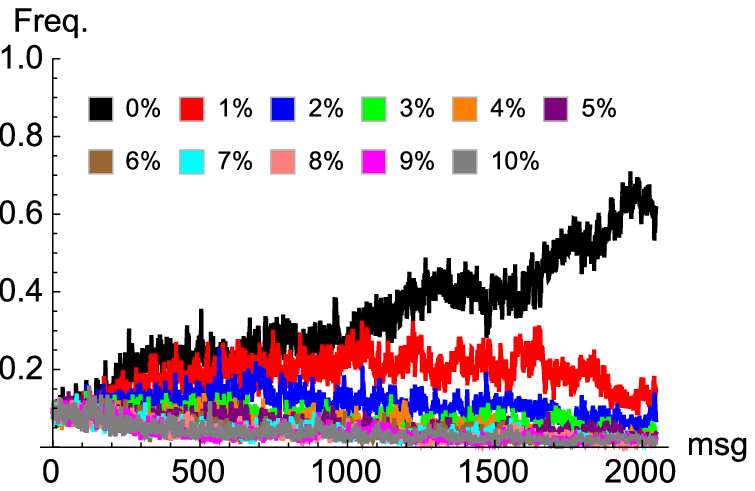}
    \label{fig:actions300Qs} }
  \caption{Actions frequencies curves. Deadline 300s, SC, $100\%$ interested nodes
    \label{fig:actions120}}
\end{figure*}

\subsubsection{Analysis of the injection policies}

In order to better understand the reason behind the  presented offloading performance,
in this section we study the temporal evolution, content after content, of the
(re-)injection actions performed by the central controller for all the
considered approaches and content delivery deadlines. Specifically, for each
simulation run we computed the  frequencies of how many times each action has
been taken by the controller. Then, we averaged results over $10$ runs.
Without loss of generality, for
this analysis we focus on the single community scenario in which all nodes are
interested in receiving the content.  In Figure \ref{fig:actions1000} we
collected the behaviour of all the approaches corresponding to scenarios with a
content deadline of $1000s$. In that experiment, both AC and Droid offloaded
$85\%$ of traffic. Interestingly, this result is reflected in the type of
actions taken by both approaches. Apparently, in this scenario the best strategy
is to perform an initial injection and then do nothing for the rest of the time.
In fact, DR does a first injection using $9\%$ of seeds and no other injections
 follow. AC shows the ability to learn that  this is the best strategy: after
the learning phase we see that the most performed action is $0\%$ of injection
and the second most taken action is $1\%$.  Qe, given that there
is enough time to let the opportunistic dissemination evolve, is able to
learn 
the best policy.  Qs, instead, has a totally different behaviour. As we can
notice  almost all the actions are taken with the same frequency and the use of
the $0\%$ action starts to increase sensitively just after 300 messages.  This
behaviour is the direct consequence of the slow learning performance of Qs.

Finally we performed the same analysis for content with $300s$ delivery
deadline. Such a tight deadline forces the approaches to find a policy with a
good balance between how aggressive  injections
should be w.r.t. the
dissemination capabilities of the opportunistic network. As reported  in Figure
\ref{fig:actions120} we see that  AC and DR achieve better  performance because
of a more efficient strategy: few  injections followed by many $0\%$
actions to permit the opportunistic dissemination do the rest. Qs, as shown in
Figure \ref{fig:actions300Qs}, although very slowly, proves to be able to learn
a strategy very similar to AC, leading its performance towards those of AC and
DR. Qe instead, has  worse performance (only $37\%$) because in the
policy it learns the $0\%$ action is performed too many times with respect to the others.

\section{Conclusion}
\label{sec:conclusions}

In this paper we propose a new
solution to reduce the load of the cellular infrastructure. Precisely, the
cellular infrastructure offloads part of the traffic to a complementary
opportunistic network formed by users' mobile devices, and a subset of
nodes is used to seed the dissemination of contents in the opportunistic network.  In
order to minimise the number of transmissions over the cellular network, we
designed a Reinforcement Learning  procedure through which the controller
learns, over time, what is the most rewarding injection policy. Precisely, our
offloading procedure is general enough to be independent of the specific
Reinforcement Learning algorithm that can be applied. To evaluate the
performance and the generality of our approach, in this paper we used two well
know learning algorithms: Actor-Critic and Q-Learning. Through simulations we
have shown that the two learning algorithms have very different performance in
all the considered scenarios, and Actor-Critic proves to be superior to
Q-Learning. Moreover our results demonstrate that our offloading mechanism
(based on Reinforcement Learning) is able to offload up to 97\% of the traffic,
and outperforms state-of-the art approaches that are not based on
  learning algorithms, although set at their best configuration
for the tested scenarios. Quite significantly, Actor-Critic is able to
adapt dynamically to radically different conditions (e.g., in terms of mobility
patterns and content delivery deadlines), while state-of-the-art approaches approaches would
need to be reconfigured for each new configuration. This guarantees a very
significant performance advantage to Actor-Critic, in the order of 20\%
additional offloaded traffic in the tested scenarios. All in all, these results
show that Reinforcement Learning algorithms (and Actor-Critic in particular) are
very suitable solutions for controlling data offloading from cellular networks,
as they are very adaptive, and able to learn the optimal policy in a range of
different configurations.
  
\section*{Acknowledgments}
This work was partially funded by the European Commission under the MOTO (FP7 317959) and EIT ICT Labs MOSES (Business Plan 2015) projects.
\section*{References}
\bibliography{bibliography}

\end{document}